# Remote Measurement of Heliostat Reflectivity with the Backward Gazing Procedure

François Hénault[1[0009-0008-4825-5519]], Gilles Flamant[2], Cyril Caliot[3]

[1] Institut de Planétologie et d'Astrophysique de Grenoble, Université Grenoble-Alpes, CNRS, B.P. 53, 38041 Grenoble, France

[2] Processes, Materials and Solar Energy laboratory, PROMES CNRS, 7 Rue du Four Solaire, 66120 Font-Romeu-Odeillo-Via, France

[3] CNRS, UPPA, E2S, LMAP, 1 Allée du parc Montaury, 64600 Anglet, France

**Abstract.** Concentrated solar power is a promising technique enabling renewable energy production with large scale solar power plants in the near future. Estimating quantitatively the reflectivity of a solar concentrator is a major issue, since it has a significant impact on the flux distribution formed on the solar receiver. Moreover, it is desirable that the mirrors can be measured during operation in order to evaluate environmental factors such as day/night thermal cycles or soiling and ageing effects at the reflective surfaces. For that purpose, we used a backward gazing method that was originally developed to measure mirror shape and misalignment errors. The method operates in quasi real-time without disturbing the heat production process. It was successfully tested at the Themis solar tower power plant in Targasonne, France. Its basic principle consists in acquiring four simultaneous images of a Sun-tracking heliostat, captured from different observation points located near the thermal receiver. The images are then processed with a minimization algorithm allowing the determination of mirror slopes errors. In this communication, it is shown that the algorithm also allows one to get quantitative reflectivity maps at the surface of the heliostat. The measurement is fully remote and is used to evaluate surface reflectivity that depends on optical coatings quality and soiling. Preliminary results obtained with a Themis heliostat are presented. They show that reflectivity measurements can be carried out within repeatability about ☐5% Peak-to-Valley (PTV) and 1% RMS. Ways to improving these numbers are discussed in the paper.

## Introduction

Concentrated solar power (CSP) is a promising technique enabling renewable energy production with large scale solar power plants in the near future. In CSP tower plants, the reflectivity of the heliostats plays a major role on the achieved performance and system efficiency, implying that the mirrors must be cleaned regularly. Thus it is highly desirable to perform measurements of the heliostats reflectivity in situ and in quasi real-time.

Heliostats reflectivity losses are known to originate from dust deposition in dessertic environment, optical coatings degradation due to day/time thermal cycles and humidity, and more generally from any damage of the optical surfaces. Ways of measuring them have been extensively reviewed in Ref. [1]. They can be schematically divided into two families:

- Using portable reflectometers such as described in Refs. [2-4]. These measurements are generally restricted to mirror samples in laboratory. Extending them to in situ mirror measurements is feasible, but would require excessive measurement time for heliostat fields comprising hundreds or thousands of mirrors, multiplied with the number of measurement points on each mirror.
- Performing remote measurements as described in Refs. [5-6] that make use of different images of the flux density formed at the solar receiver. They allow estimating the global reflectivity loss of the heliostats, but provide no information about their locations and amplitudes.

Here is described a local, backward gazing method originally developed to measure the mirror shape and misalignment errors of the heliostats in a reasonable period of time. It allows quasi real-time measurements without disturbing the heat production process. Its basic principle consists in acquiring four simultaneous images of a Sun-tracking heliostat, captured from different observation points located near the solar receiver. The images are then processed with a minimization algorithm allowing the determination of mirror slopes errors. In this communication, it is shown that the algorithm also allows one to get a quantitative reflectivity map at the surface of the heliostat. The measurement is fully remote and allows evaluating soiling effects due to dust accumulation and moisture, as well as surface defects and cracks inside the optical coatings.

The paper is divided as follows: section 2 firstly describes the Themis experiment and the reflectivity reconstruction algorithm. The measurement methodology and the obtained numerical results are given in section 3, and then discussed in section 4. A brief conclusion is drawn in section 5.

## Method

The backward gazing method was developed in the 1980's to measure the canting and shape errors of the reflective facets of solar concentrators, such as those equipping the 1 MW solar furnace in Odeillo, France, and the focusing heliostats of the solar tower power plant Themis in Targasonne, France [7]. Later, the appearance of modern CCD cameras allowed using more than one single point of observation, therefore achieving quantitative measurements and reinforcing considerably the interest of the method. Numerical simulations were undertaken in order to evaluating its performance, and demonstrated that a measurement accuracy of the mirror slopes and misalignment errors better than 0.1 mrad is feasible [8-12]. A series of experiments were then conducted at the Themis solar power plant, and confirmed a high potential of the method for measuring the slopes errors of the heliostats [13]. All data acquired during these experiments are reusable for quantitatively estimating the reflectivity maps of the heliostats.

### Coordinate systems and scientific notations

The Themis experiment is illustrated in Figure 1. It makes use of the following coordinate systems (Figure 1-a):

- The XYZ reference frame is attached to an individual heliostat with X its optical axis and YZ are its lateral dimensions along which its geometry is defined (see Figure 1-d). Points at the surface of the heliostat are denoted P(y, z) with y and z their Cartesian coordinates.
- The X'Y'Z' reference frame is attached to the solar receiver, or to the target plane. The X'-axis is directed from the centre of the heliostat to the centre of the target plane. The Y' and Z' axes are assumed to be perpendicular to the X'-axis. The four cameras are installed at points M'i (1 ≤ i ≤ 4) of Cartesian coordinates (y'i, z'i).

In addition, three vectors defined:

- **S** is a unitary vector directed to the Sun centre,
- **N** is a unitary vector perpendicular to the heliostat surface, parallel to the X-axis,
- **R** is the unitary target vector parallel to the X'-axis.

The vectors **S**, **R** and **N** obey the Smells-Descartes reflection law that writes in vectorial form:

$$\mathbf{S} + \mathbf{R} = 2(\mathbf{SN})\mathbf{N}. \tag{1}$$

The different input and output maps at the heliostat surface employed here are summarized in Table 1. It may be noted that theoretical and approximated relations between the angles $\varepsilon_i(P)$, $a(P)$ and $h(P)$ were established in Refs. [8-12]. They are not utilized here because the minimization algorithm described in the next subsection allows removing any approximation.

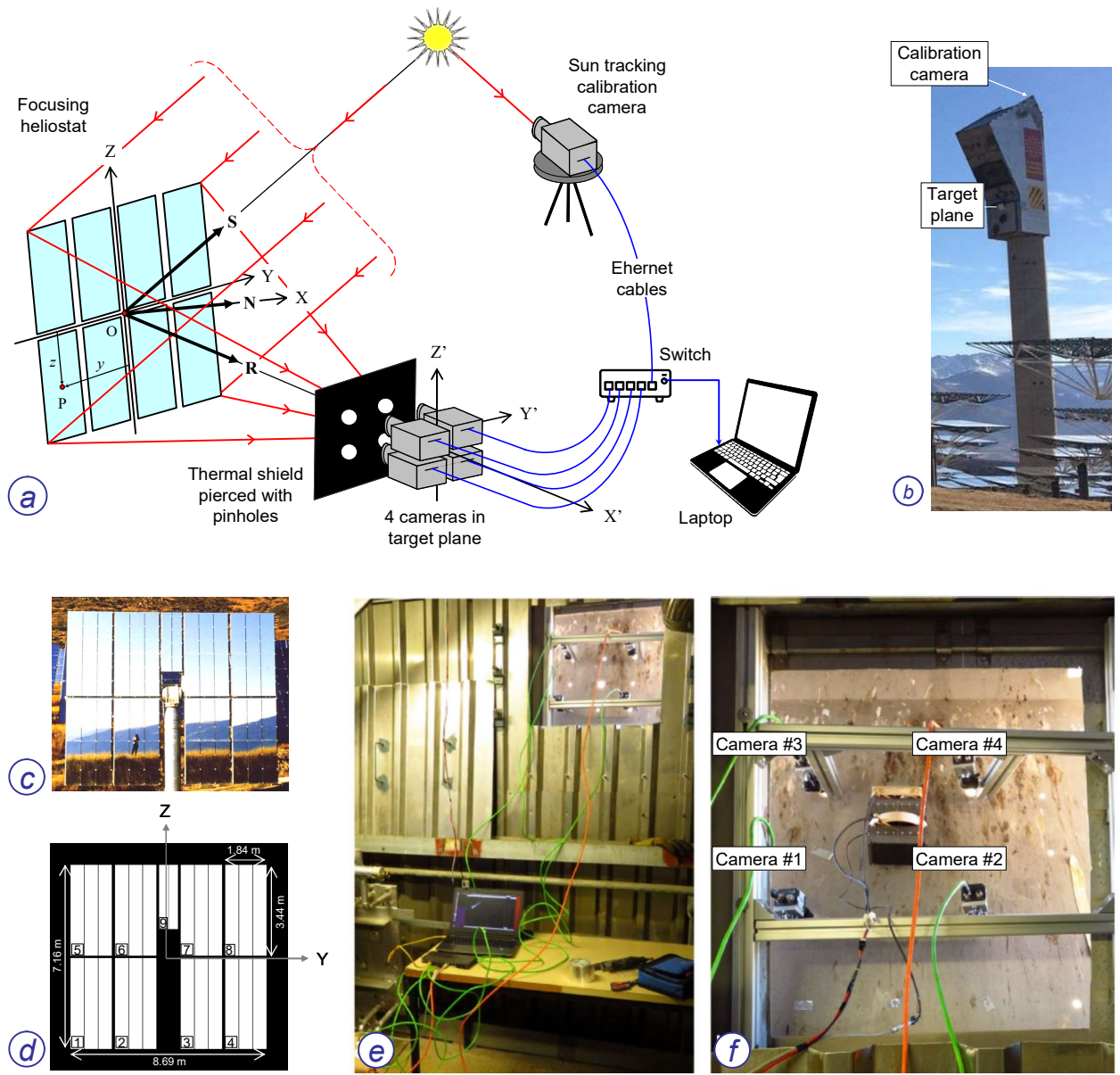

**Figure 1**: Principle of the four cameras backward gazing method and its implementation at the Themis solar power plant.

Table 1: Input and output maps at the surface of the heliostat.

| Input and output maps | Symbol | Unit |
|---|---|---|
| Image acquired with the $i^{th}$ camera ($1 \leq i \leq 4$) | $H_i(P)$ | – |
| Simulated image for the $i^{th}$ camera ($1 \leq i \leq 4$) | $B_i(P)$ | – |
| Deviation angle with respect to Sun centre for the $i^{th}$ camera ($1 \leq i \leq 4$) | $\varepsilon_i(P)$ | mrad |
| Heliostat reflectivity map | $R(P)$ | – |
| Heliostat slope errors map in azimuth along Y-axis | $a(P)$ | mrad |
| Heliostat Slope errors map in height along Z-axis | $h(P)$ | mrad |

## The Themis experiment

The Themis experiment was extensively described in Refs. [11] and [13]. Its main features are illustrated in Figure 1 and summarized below.

- The measured heliostat is made of nine focusing modules, eight of them being strictly identical. A 9[th] "complementary" module is located just above the rotating elevation mechanism (see Figures 1-c and 1-d). The modules are tilted one with respect to the other in order to mimic an ideal parabolic profile. The overall dimensions of the heliostat are 8.75 x 7.34 m along the Y and Z axes respectively. The heliostat is located at a distance $d$ = 131 m from the target plane and is set in Sun-tracking mode.
- Four small cameras equipped with CMOS monochrome sensors and telephoto lenses are used to capturing images of the Sun reflected through the heliostat with a maximal resolution of 1280x1024 pixels. They are located behind a thermal shield pierced with four 25-mm diameter pinholes in the Y'Z' target plane, enabling the observation of the heliostat field. They are protected from the concentrated solar radiation by a set of neutral densities. The distance between the cameras is set to 200 mm (see Figures 1-e and 1-f). The common acquisition time of all mages is set to 2 milliseconds, which is negligible with respect to the Sun tracking refreshing rate of he heliostat drive.
- A fifth CMOS camera is located at the top of the solar tower (see Figure 1-b) and mounted on a Sun-tracking mechanism. It is used for radiometric calibration of the images acquired with the four previous cameras.
- Data from the five cameras are acquired simultaneously and transferred to a laptop computer via Ethernet cables and a switch.

The image data processing software is then executed offline, and is described in the next subsection. It may be noted that a similar experimental setup was described in Ref. [14] in order to estimating the angular deviations of Sun-tracking heliostats. However the acquired data was not utilized in view of reflectivity measurements.

## Reflectivity reconstruction algorithm

Then, starting from the four pre-processed camera images $H_i$ (P), the reflectivity reconstruction algorithm consists in the following steps (see Figure 2):

1. Select a point P at the surface of the heliostat.
2. Read the brightness value at point P $H_i$ (P) from the image of the heliostat recorded with the $i^{th}$ camera ($1 \leq i \leq 4$).
3. Perform reverse ray-tracing starting from point M'$_i$, then reflecting the ray at point P and finally directing it to the solar disk.
4. Compute the angular deviation $\varepsilon_i$ (P) of the reverse reflected ray with respect to the Sun centre.
5. Estimate the image brightness $B_i$ (P) at point P, either from an analytical model of from the direct Sun image recorded with the calibration camera. In the first option we use Jose's formula [15]:

$$B_i(P) = 0.39 + 0.61 \sqrt{1 - \frac{\sin^2 \varepsilon_i(P)}{\sin^2 \varepsilon_0}} ,\qquad(2)$$

   with $\varepsilon_0$ the angular radius of the Sun taken equal to 16 arcmin.
6. Repeat steps 1 to 5 with another camera $i' \neq i$.
7. Compute a cost function defined as:

$$CF = \sqrt{\sum_{i=1}^{4} (H_i(P) - R(P) \times B_i(P))^2} .\qquad(3)$$

8. Find the minimum value of the cost function CF when varying the reflectivity factor R(P) and the deviation angles $a$(P) and $h$(P) with a Powell descent algorithm.
9. Repeat steps 1 to 8 for all points P at the heliostat surface.

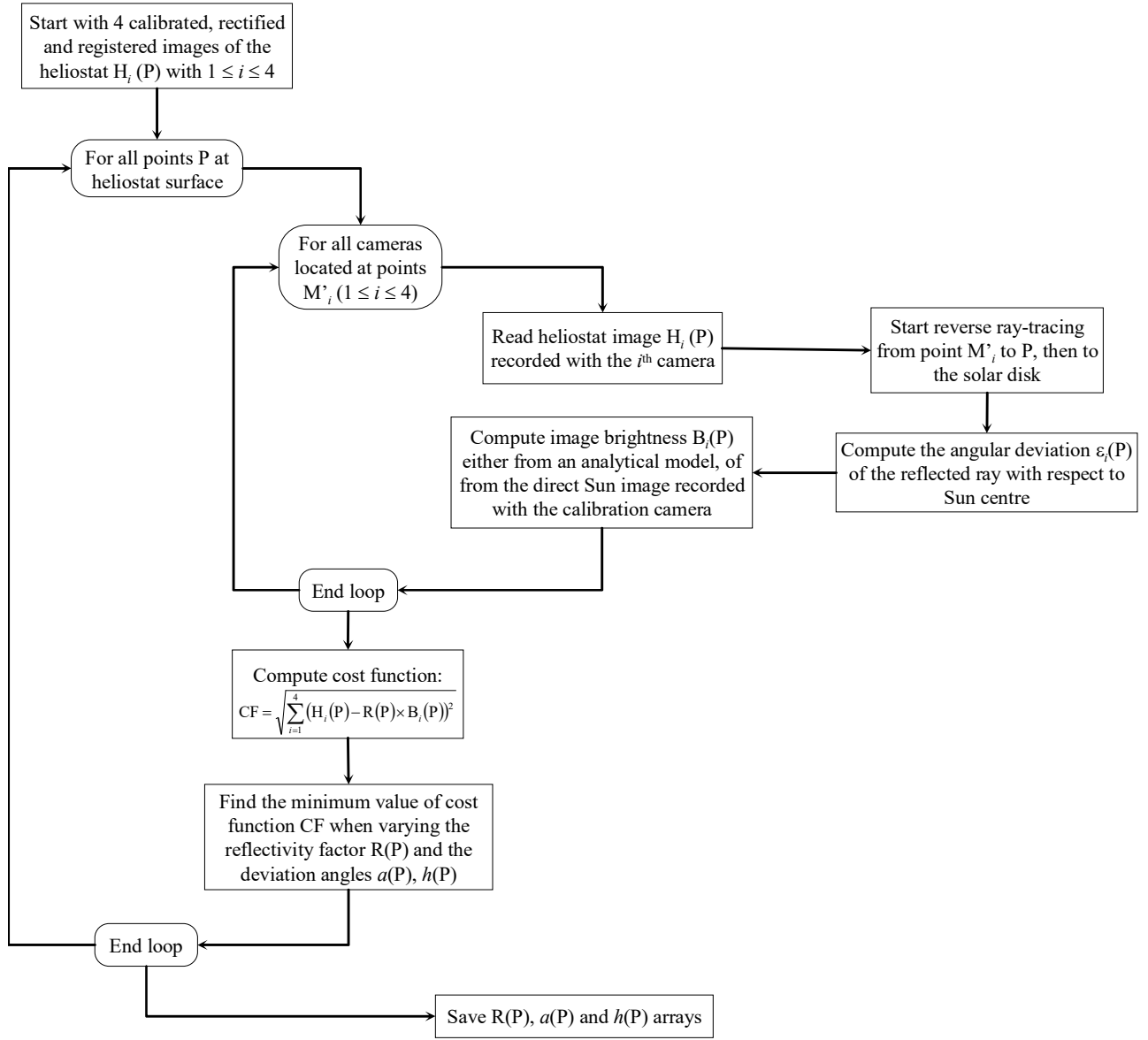

**Figure 2**: Flow-chart of the reflectivity and slopes reconstruction algorithm.

## Methodology and measurement results

Mapping the full reflectivity distribution R(P) of a heliostat and estimating the global measurement accuracy faces up to a serious difficulty that is the absence of reference measurements to be compared with: in that case, only portable reflectometers could be used [2-4] at the price of excessive measuring time and significantly reduced spatial sampling. Thus we opted for the following methodology:

- A first set of measurements was carried out on the 21$^{st}$ of December 2017 at 14h11 GMT, which corresponds to solar angles $a_S$ and $h_S$ equal to -34.3 deg. in azimuth and +19.9 deg. in elevation.
- A second set of measurements was acquired 30 minutes later with solar angles $a_S$ = -36.2 deg. and $h_S$ = +15.2 deg.

- It is assumed that reflectivity changes due to the slight variations of the incidence angles on the heliostat (< 3 deg.) are negligible, and that other reflectivity degradations do not occur in this short lap of time.
- Then the difference of reflectivity measurements between case 1 and 2 stands for a fair estimator of the repeatability error.
- It is finally assumed that the absolute measurement accuracy and repeatability are of the same magnitude order.

The acquired heliostat images for both cases n°1 and 2 are reproduced in Figure 3. The measured repeatability is given in Table 2 and illustrated with the false-colour views in Figure 4. The Table 2 shows the spatial coverage on each heliostat module, which is proportional to the number of valid pixels where the cost function does not exceed a certain threshold. The average, PTV and RMS reflectivity errors are indicated in the rightmost columns for each heliostat module. It must be noticed that the module n°1 was excluded from these statistics because of a too low spatial coverage. The estimated repeatability is then found to be lower or equal than ☐ 5% PTV and 1% in RMS sense. These results are further discussed into the next section.

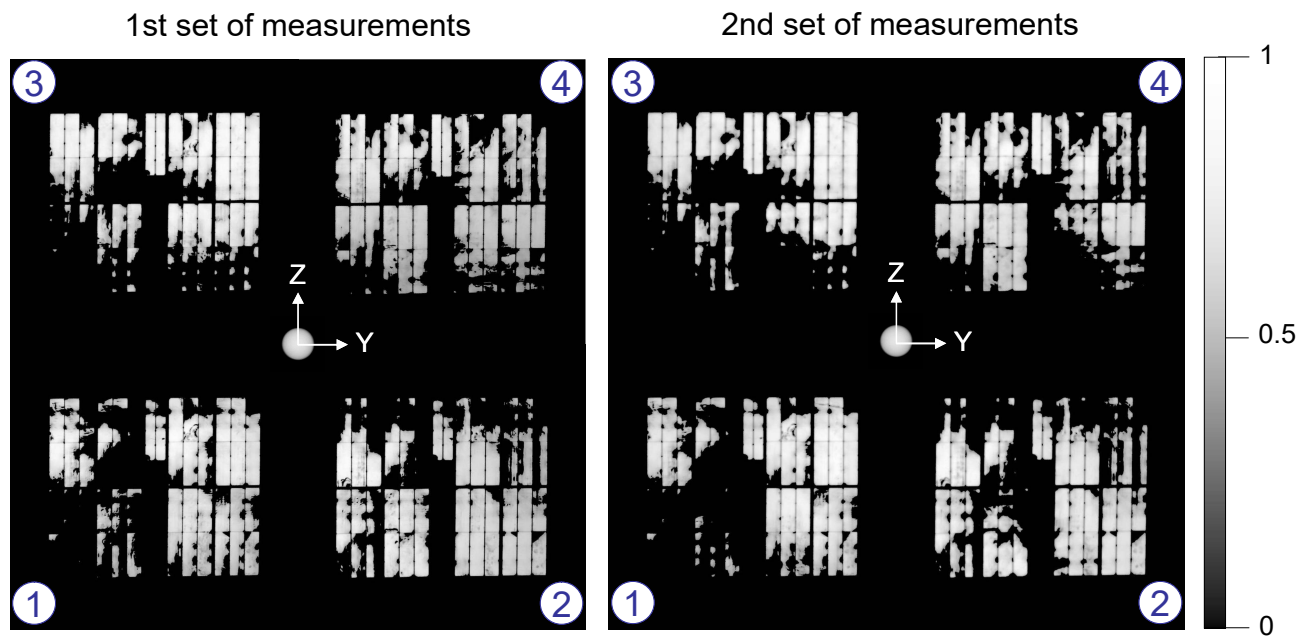

**Figure 3**: Acquired images during the first and second sets of measurements.
Direct Sun images acquired with the fifth camera are displayed at the centre.

**Table 2**: Estimation of reflectivity measurement errors for each heliostat module.

| Module number | Spatial coverage (%) | Reflectivity measurement error | | |
|---|---|---|---|---|
| | | Mean (%) | PTV (%) | RMS (%) |
| 1 | 0,3 | 1,0 | ± 1,8 | 1,0 |
| 2 | 5,0 | 0,3 | ± 4,8 | 0,9 |
| 3 | 24,3 | 1,7 | ± 5,6 | 1,2 |
| 4 | 36,2 | 2,5 | ± 4,3 | 1,4 |
| 5 | 38,9 | -0,7 | ± 5,5 | 1,4 |
| 6 | 8,6 | -0,6 | ± 4,8 | 1,3 |
| 7 | 31,0 | 0,6 | ± 5,0 | 1,6 |
| 8 | 22,7 | 0,8 | ± 5,1 | 0,9 |
| 9 | 63,9 | 0,2 | ± 5,0 | 0,8 |
| Average | | 0,6 | ± 4,7 | 1,2 |

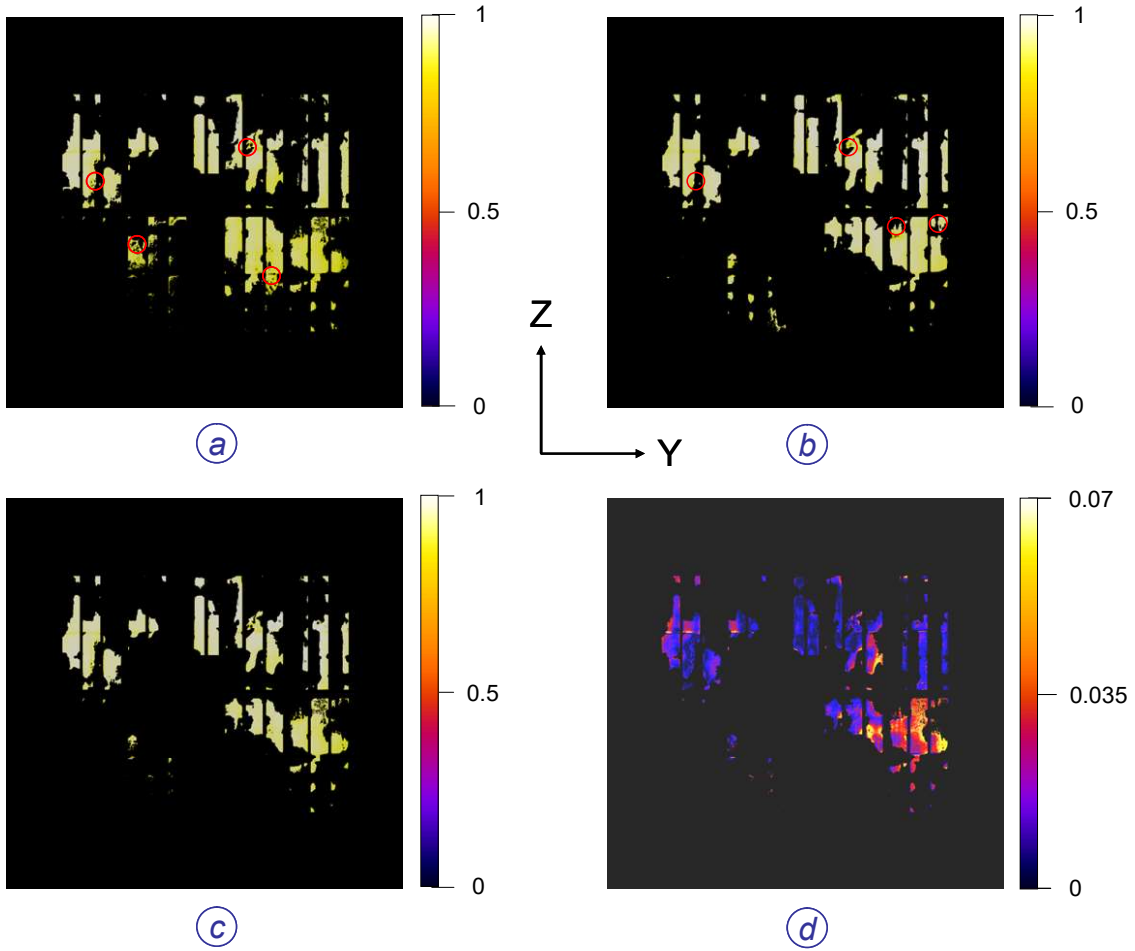

**Figure 4**: (a) Initial measurement. (b) Second measurement performed 30 minutes later. (c) Averaged reflectivity map. (d) Estimation of the repeatability error between both measurements. The maximal values of the images are normalized to unity. Red circles indicate probable cracks locations.

# Discussion

The most important issue probably consists in extending the spatial coverage of the method in order to reconstruct the entire surface of the heliostat. Actually, the limited spatial coverage seen in Figure 4 results from the combination of large slope errors $a$(P) and $h$(P) with a limited number of cameras $N_{Cam}$ = 4. It was demonstrated in Ref. [16] that using more cameras, either arranged in a square or a line geometry combined with sun-tracking operation allows overcoming this difficulty, and would improve the measurement accuracy by a by a factor $1/\sqrt{N_{Cam}}$ at the same time.

Secondly, careful examination of the reflectivity maps in Figure 4 allows identifying the areas where the measurement accuracy is degraded. It is found that they coincide with those areas where the deviation angles $a$(P) and $h$(P) are poorly determined [13]. The degraded areas are generally located near the contours of the reflectivity maps where steep transitions between the lighted and unlighted zones are observed.

Moreover, the method also allows to locating precisely the corrupted areas. In particular, cracks in the optical coatings can be evidenced. Some of them are marked with red circles in Figure 4. It turns out that this will be very helpful in deciding if and when cleaning or replacing some heliostat modules with spare ones is required.

Lastly, an extensive analysis of the main experimental error sources was presented in Ref. [13]. Ways of mitigation were proposed as follows:

For each camera, increasing the number $N_i$ of acquired images during the refreshing time of the tracking heliostat that is currently limited to $N_i$ = 5. A reduction of the noise by a factor $1/\sqrt{N_i}$ will result.

Implementing real-time visualization of the images of the observed heliostat, assisted by remote controlled zoom and focusing devices of the cameras.

Improving the registration algorithm of the rectified images of the heliostat down to sub-pixel level.

Provided that such improvements are implemented, we believe that a measurement accuracy of ≈ 2% PTV and 0.5% RMS is achievable on the full heliostat surface.

# Conclusion

Estimating quantitatively the reflectivity of solar concentrators will be a key issue for increasing the energy produced by large scale solar power plants in the near future. Here was described the principle of a backward gazing method originally intended to measure the shape and canting errors of focusing heliostats. The method proves to be very efficient for regular control of the heliostats reflectivity with a high spatial resolution. It enables precise estimation of various environmental factors such as day/night thermal cycles, soiling and ageing effects on the reflective surfaces due to dust accumulation and moisture, or of cracks in the optical surfaces. It is fully remote and can be operated in quasi real-time when the heliostats are in Sun-tracking mode, without disturbing the electricity production process. A minimization algorithm then allows determining quantitative reflectivity maps at the surface of the heliostats. Preliminary results obtained with a focusing heliostat of the Themis solar tower power plant showed that the current measurement errors are about □5% PTV and 1% RMS. Ways to improving these numbers were discussed and may allow attaining an accuracy of □2% PTV and 0.5% RMS. The method may finally be used for routine reflectivity measurements helping in deciding if and when some heliostat modules have to be cleaned or replaced.

# Author contributions

F. Hénault is the First Author. He is optical engineer, PHD in Optics and Photonics and acquired extensive knowledge about the opto-mechanical design of focusing heliostats.